\documentclass[namedreferences]{solarphysics}
\usepackage[optionalrh]{spr-sola-addons} 
\usepackage{epsfig}          
\usepackage{graphicx}        
\usepackage{color}           
\usepackage{url}             
\usepackage{lscape}
\usepackage{longtable}
\usepackage{multirow}

\usepackage{amsmath}
\usepackage{amssymb}
\usepackage{longtable}
\usepackage{array}

\usepackage{color}
\usepackage{multirow}
\usepackage{lscape}

\newcommand{\etal}{{\it et al.}}

\newcommand{\aap}{    {\it Astron. Astrophys.}}

\newcommand{\apj}{    {\it Astrophys. J.}}
\newcommand{\apjl}{    {\it Astrophys. J. Lett.}}

\newcommand{\jgr}{    {\it J. Geophys. Res.}}

\newcommand{\nat}{    {\it Nature}}

\newcommand{\solphys}{{\it Solar Phys.}}

\begin{document}
\begin{article}
\begin{opening}
\title{Energy Build-up and Triggering Leading to a M1.5 Flare on 1 August 2014}
\author{S. Liu$^{1,2}$  \sep J.T. Su$^{1,2}$}
\runningauthor{S. Liu, J.T. Su}
\runningtitle{Energy Build-up and Triggering Leading to a M1.5 Flare}
\institute{$^{1}$Key Laboratory of Solar Activity, National
Astronomical Observatory, Chinese Academy of Sciences, Beijing, China
$^{2}$School of Astronomy and Space Sciences, University of Chinese Academy of Sciences, Beijing, China \\
        email: \url{lius@nao.cas.cn}\\}
\begin{abstract}

The energy storage and trigger mechanisms of solar flares are important for understanding of solar activity. We analyzed multi-wavelength observations of a M1.5 flare on 1 August 2014, in active region NOAA 12127 (SOL2014-08-01T18:13). There are evident large scale sunspot rotations in positive magnetic field of the main energy release region before the eruption; the rotations contain both clockwise and counter-clockwise directions.  Injection of magnetic helicity from the photosphere prior to the flare.  The sign of the helicity injection is reversed after the flare. It is found that both persistent larger scale ($\approx$ one day) and impulse smaller scale ($\approx$ one to two hours) magnetic-flux emergences are associated with the flare. We conclude that larger-scale flux emergence, helicity injection and sunspot rotation contribute to the energy build up, while the small-scale magnetic-flux emergence plays crucial role in triggering the flare.
\end{abstract}
\end{opening}

\section{Introduction}
Solar flare are the most common and dynamic phenomenon in the Sun,
it can give the direct and strong disturbances to space weather.
For example a major solar flare, it usually releases more than
10$^{32}$ ergs and ejects more than 10$^{16}$ g plasma into interplanetary space.
Traditionally, a solar flare is regarded as the product of conversions from magnetic energy stored in a active region
to heat and kinetic energy carried by erupted plasma.
A typical flare emission can be exhibited nearly at all the solar atmosphere,
from the photosphere/chromosphere to the transition region and corona.

Solar flares basically arise from the disruptions of active-region magnetic field,
and the mechanism of magnetic reconnections probably plays a key role during this process
(\citeauthor{1957JGR....62..509P}, \citeyear{1957JGR....62..509P};
\citeauthor{2000mrmt.conf.....P}, \citeyear{2000mrmt.conf.....P};
\citeauthor{2003NewAR..47...53L}, \citeyear{2003NewAR..47...53L}).
When magnetic reconnections take place, the magnetic-field lines with oppositely orientations mutually annihilate
at the same time that results in the conversion from magnetic energy
to heat and kinetic energy carried by magnetized plasma spontaneously.
As for magnetic reconnection, the previous standard model is 2D CSHKP model
(\citeauthor{1964NASSP..50..451C}, \citeyear{1964NASSP..50..451C};
\citeauthor{1966Natur.211..695S}, \citeyear{1966Natur.211..695S};
\citeauthor{1974SoPh...34..323H}, \citeyear{1974SoPh...34..323H};
\citeauthor{1976SoPh...50...85K}, \citeyear{1976SoPh...50...85K}).
In CSHKP mode, the magnetic reconnections occur at magnetic null-point
(locally zero magnetic field), where magnetic field lines with opposite directions
meet. This model can explain many flare properties, but for some 3D properties it fails.
Later, advanced breakout and tether-cutting models are given more attention
(\citeauthor{1999ApJ...510..485A}, \citeyear{1999ApJ...510..485A};
\citeauthor{2001ApJ...552..833M}, \citeyear{2001ApJ...552..833M};
\citeauthor{2006SoPh..235..147Y}, \citeyear{2006SoPh..235..147Y};
\citeauthor{2008ApJ...689L.157Z}, \citeyear{2008ApJ...689L.157Z};
\citeauthor{2010ApJ...721.1579R}, \citeyear{2010ApJ...721.1579R};
\citeauthor{2014SoPh..289.2091L}, \citeyear{2014SoPh..289.2091L}).
These two models are related to catastrophic instability, for tether cutting the reconnections occur at the lower current
sheet, while for breakout they occur at higher current sheet (\citeauthor{2014SoPh..289.2091L}, \citeyear{2014SoPh..289.2091L}).

The trigger mechanisms of flare are important aspects that worth to be
studied, and they are open problems that
have not been completely and clearly resolved. Parts of flares can be triggered through magnetic-flux emergences,
the magnetic reconnections occur between the pre-existing flux and the emerging flux when they
favor facilitating the magnetic conditions for reconnections
(\citeauthor{1993A&A...271..292D}, \citeyear{1993A&A...271..292D};
\citeauthor{2012ApJ...760...31K}, \citeyear{2012ApJ...760...31K};
\citeauthor{2014ApJ...796...44K}, \citeyear{2014ApJ...796...44K};
\citeauthor{2015SoPh..290.3641L}, \citeyear{2015SoPh..290.3641L}).
During magnetic mergence, the magnetic-flux ropes are formed and detected.
Additionally, in some cases of magnetic emergences the flux ropes formations
are especially important conditions for flare erupting
(\citeauthor{2012NatCo...3E.747Z}, \citeyear{2012NatCo...3E.747Z};
\citeauthor{2015NatCo...6E7008W}, \citeyear{2015NatCo...6E7008W}).
As well as the cancellations of flux can also lead to flares,
which can regarded as an alternative mechanism for initiating flares.
(\citeauthor{1989SoPh..121..197L}, \citeyear{1989SoPh..121..197L};
\citeauthor{2010A&A...521A..49S}, \citeyear{2010A&A...521A..49S};
\citeauthor{2013SoPh..283..429B}, \citeyear{2013SoPh..283..429B}).
As the processes of magnetic cancellations at the photosphere where coronal ropes rooted,
the reconnections happen consequently at low atmosphere.
As a result a flux rope above magnetic neutral lines and a filament channel in the photosphere and chromosphere
gradually formed where magnetic cancellations occur, when the formed flux ropes become
unstable then the flares erupt
(\citeauthor{1993SoPh..143..119W}, \citeyear{1993SoPh..143..119W};
\citeauthor{2010ApJ...708..314A}, \citeyear{2010ApJ...708..314A};
\citeauthor{2011ApJ...742L..27A}, \citeyear{2011ApJ...742L..27A};
\citeauthor{2000JGR...105.2375L}, \citeyear{2000JGR...105.2375L}).
Another  physical quantity that worth to pay attention is helicit. The magnetic helicity, which can  characterize
the topologies of active region magnetic field, probable contains an important clue to the production of solar flare
(\citeauthor{2006ApJ...644..575Z}, \citeyear{2006ApJ...644..575Z};
\citeauthor{2006SoPh..234...21L}, \citeyear{2006SoPh..234...21L};
\citeauthor{2008ApJ...686.1397P}, \citeyear{2008ApJ...686.1397P}).
The extent and trend of  helicity accummulations for a special active region maybe an importan reference for flare predictions.
For example, the threshold of helicity may exist for an active region to give the eruption of flare (\citeauthor{2006ApJ...644..575Z}, \citeyear{2006ApJ...644..575Z}), and the sign changes of  helicity transport maybe also relate to the production of solar flare
(\citeauthor{2008ApJ...686.1397P}, \citeyear{2008ApJ...686.1397P}).

In this article, the eruption of a M1.5 flare is displayed, which is probably triggered by
remote and complex magnetic emergences of small-scale magnetic structures.
The properties of the main flare region and trigger region are exhibited through multi-bands observations
and magnetic-field observations based on \emph{Atmospheric Imaging Assembly}
(AIA: \citeauthor{2012SoPh..275...17L}, \citeyear{2012SoPh..275...17L})
 and \emph{Helioseismic and Magnetic Imager} (HMI: \citeauthor{2012SoPh..275..229S}, \citeyear{2012SoPh..275..229S})
instruments onboard Solar Dynamics Observatory (SDO, \citeauthor{2012SoPh..275....3P}, \citeyear{2012SoPh..275....3P}).
The article is organised as follows. The observations and data processing are described in Section 2. The
results obtained are in Section 3. Finally, in Section 4, the brief conclusions and discussions are
given.
\section{Observation Data}
AIA and HMI are two main instruments onboard SDO launched in 2010.
AIA takes 4k $\times$ 4k full-disk images of the Sun in three UV-visible  and seven EUV channels with a resolution of 0.6$^{\prime\prime}$ pixel$^{-1}$. HMI obtains full-disk magnetograms with 4k $\times$ 4k CCD
in the photospheric absorption line Fe {\sc i} centered at the wavelength 6173.3~\AA{} with high spatial and temporal
resolutions of 0.6$^{\prime\prime}$ pixel$^{-1}$ and 45 s, respectively. In this study, the channels of 304~\AA{}, 171~\AA{}, and 1600~\AA{}
by AIA and line-of-sight (LOS), vector magnetograms and continuum intensity by HMI are used
to analyse the process of this M1.5 flare that erupts on 01 August 2014.
The data processing is based on the standard Solar Softwares (SSW) related to these
instruments (such as \textsf{hmi$_{-}$prep.pro}, \textsf{aia$_{-}$prep.pro} and \textsf{drot.pro}). To accurately determinate the time of flare and compare multi bands observation, the X-ray flux data observed by a \emph{Geostationary Operational Environmental Satellite} (GOES) is also employed in this study.

\begin{figure}
   \centerline{\includegraphics[width=1\textwidth,clip=]{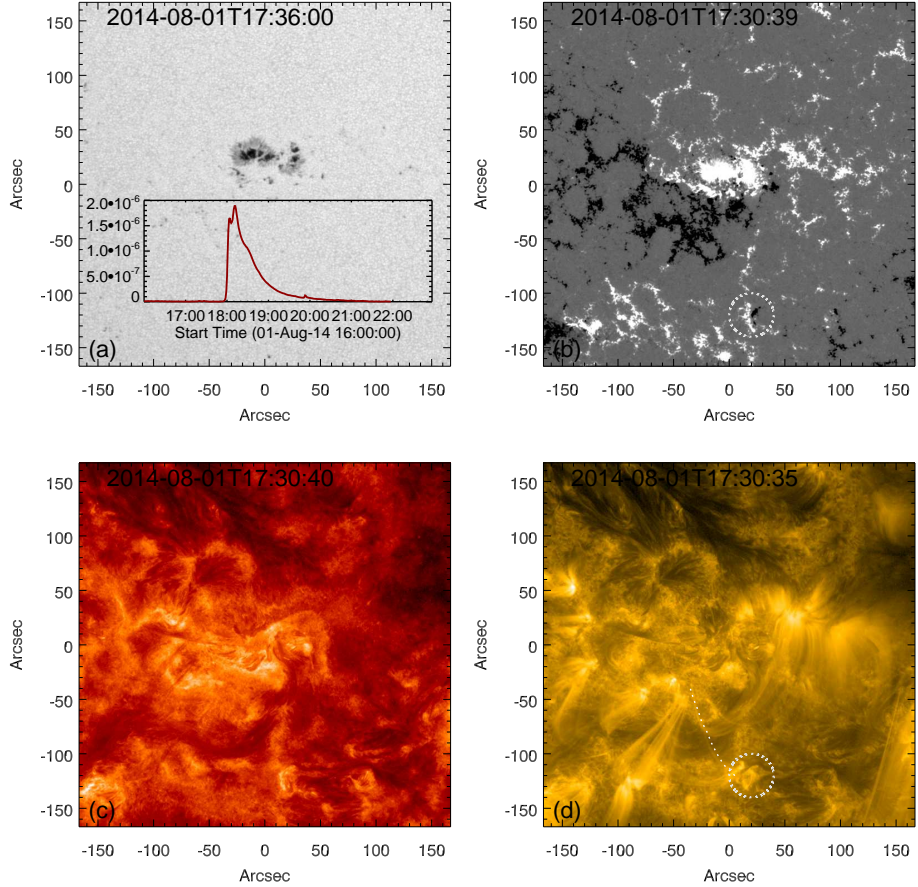}}
   \caption{The active region is shown through continuum (a), magnetogram (b), 171~\AA (c), and 304~\AA (d), and GOES curve is added in (a) to show the flare.
   The coronal loops (depicted by white dotted curve) rooted at  \emph{x:} 20 and \emph{y:} -120 connect the main region and a small-scale magnetic structure (indicated by a circle in panel (b) magnetogram) can be seen for 171~\AA~image, which maybe trigger this M1.5 flare.} \label{Fig1}
\end{figure}

The flare studied in this work erupts from NOAA 12127 active region when the active region locates at about S09W68, the flare starts at 17:55:00 UT and
stops at 18:48:00 UT with peak at 18:13:00 UT on 01 August 2014  (SOL2014-08-01T18:13). The negative front of this flare and fine structures of this active region have been studied
recently (\citeauthor{2016ApJ...819...89X}, \citeyear{2016ApJ...819...89X};
\citeauthor{2016ApJ...817..117S}, \citeyear{2016ApJ...817..117S};
\citeauthor{2016ApJ...816...30S}, \citeyear{2016ApJ...816...30S}).
This event can be regarded as a high quality candidate to study the process of flare eruption, since the last flare with level C1.3 before this M1.5 flare occur at at 02:04 on July 31.
It means the time interval is about $\approx$~40 hours, so the multiple flares effects are avoided for this M1.5 flare event analysis.
Figure \ref{Fig1} shows the active region by continuum (a), LOS magnetic field (b), 304~\AA~(c), and 171~\AA~(d) observations,
and GOES X-ray flux is over-plotted in panel a to exhibit the flare processes, which shows the pulse rise and slow drop by flux curve. The active region
displays an evident $\beta\gamma/\beta\gamma$ style, and it has relative complex connectivity.
Especially from 171~\AA, a coronal loop that connects active main region and small-scale magnetic structure located at about the position of \emph{x}=20 and \emph{y}=-120 can be seen, where the loop and small-scale magnetic structure are indicated by a circle in magnetogram map and white dotted line in the 171~\AA~image, respectively. This long coronal loop probably play an important role to triggering this M1.5 flare.

\section{Results}
\subsection{The Process of Flare Trigger}
\begin{figure}
   \centerline{\includegraphics[width=1\textwidth,clip=]{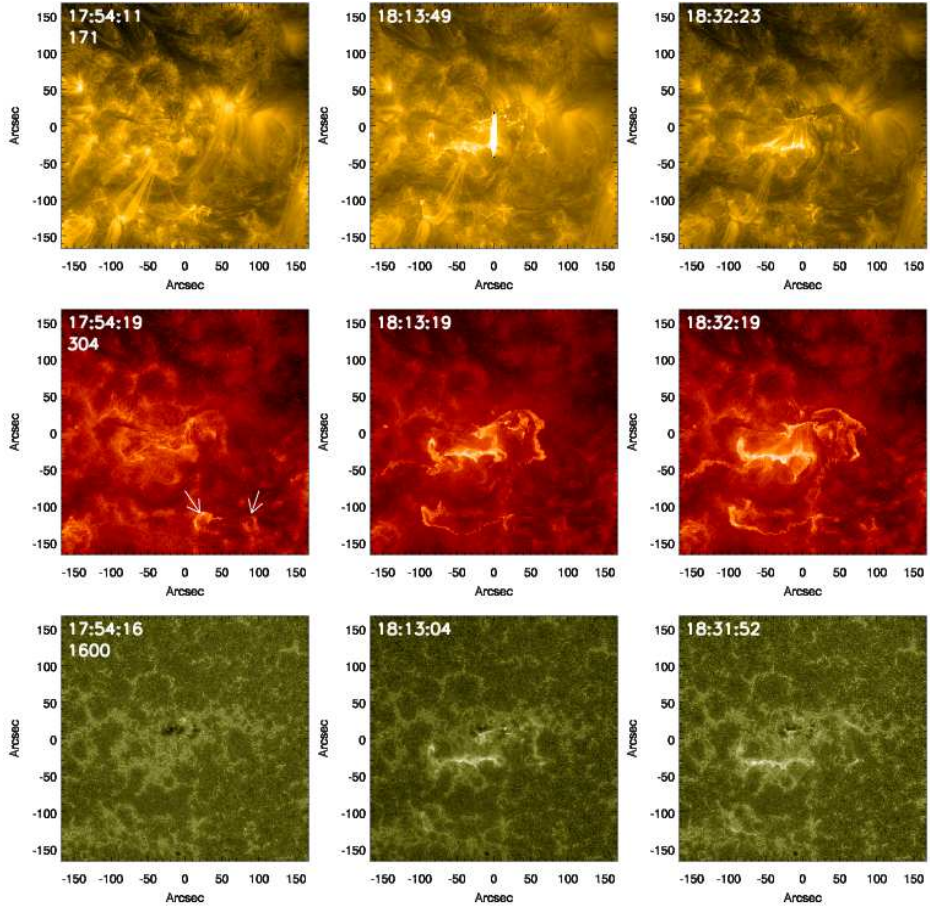}}
   \caption{Flare processes are shown by 171~\AA, 304~\AA, and 1600~\AA~pass bands observations. The left/middle/right panels are images before/during/after the flare eruption to describe
   the process of the flare. Two arrows in middle-left panel 304~\AA ~indicate the positions of the bright enhancement that maybe trigger this flare process.} \label{Fig2}
\end{figure}

Figure \ref{Fig2} shows the flare process using 171~\AA, 304~\AA, 1600~\AA~observations.
The flare ribbons can be clearly seen in these multi-bands observations around its peak time at 18:13 UT
and the post-flare loops are formed in late phase.
Additionally, multiple ribbons can be recognized that form approximatively a circular ribbon in the 304~\AA~image.
When high temporal resolution data are surveyed by time series observations AIA, especially 304~\AA~images,
two distinct bright structural features with individual and mutual dynamic characteristics are identified below the main flare region
(position 1: \emph{x}=20, \emph{y}=-120 and position 2: \emph{x}=90, \emph{y}=-120) indicate by two white arrows in the middle-left panel 304~\AA~image, and  their bright enhancements are slightly before flare in time.
Based on the LOS magnetic field shown in Figure \ref{Fig1}, it is found that the bright enhancements pointed to by white  arrows correspond to small magnetic structure.
Through time series observations data, the mass flows between these two small magnetic structures are detected before the main flare,
the structures of those mass flows also can be seen statically in Figure \ref{Fig2} between two white  arrows.

\begin{figure}
   \centerline{\includegraphics[width=1\textwidth,clip=]{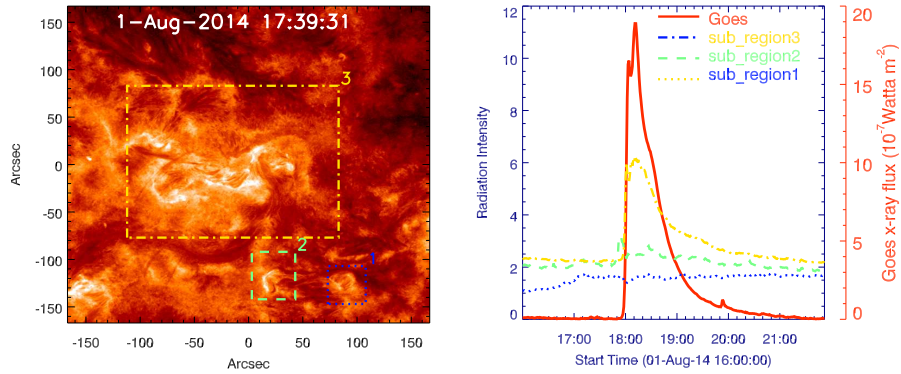}}
    \caption{Left :304~\AA~image with subregions labeled 1, 2, 3 using different color and style rectangles and numbers to show their position. Right: the evolutions of radiation intensity in individual subregion labeled by corresponding color and style lines, and the red curve shows the GOES  x-ray flux to exhibit the flare. The time difference between the max intensity of subregion 2 and subregion 3 (GOES x-ray flux) is 6.8 minutes, hence the bright enhancement of subregion 2 can be regard as the probable precursor of this flare. } \label{Fig3}
\end{figure}
In Figure \ref{Fig3}, the evolution of three evident bright enhancements regions are shown to
reveal the flare process. Here individual subregion labeled by different color numbers and rectangles in the 304~\AA~image, and the evolution of subregion are shown by the corresponding colored curves in the right panel. Subregion 1 and 2 correspond two small-scale magnetic emergences, subregion 3 is the flare main region,
in the figure the GOES x-ray flux is over-plotted on a red curve. From Figure \ref{Fig3}, it can be found that there exists a time delay
the maximum of bright enhancements between subregions 2 and 3, this time delay is 6.8 minutes.
It means that the bright enhancements of subregion 2 probably lead to the subsequent flare eruption.
The subregion 3 is the main region of the flare, its maximum of bright enhancements can be regarded as the peak of flare,
so intensity evolution of subregion 3 matches that of the GOES x-ray flux.
As for subregion 1, there is no evident pulsed bright enhancements, but its intensity increase before the flare to some extent.
Additionally subregion 1 it displays more dynamic and eject mass to subregion 2 intermittently before the bright enhancements of
subregion 2 and also flare, which all mean that there are complex relations between these two subregions.
The whole picture of this flare process can be depicted that the interactions with mass flows and bright enhancements between subregion 1 and 2
lead to the bright enhancements of subregion 2, then trigger the main region to produce the flare.
\begin{figure}
   \centerline{\includegraphics[width=1\textwidth,clip=]{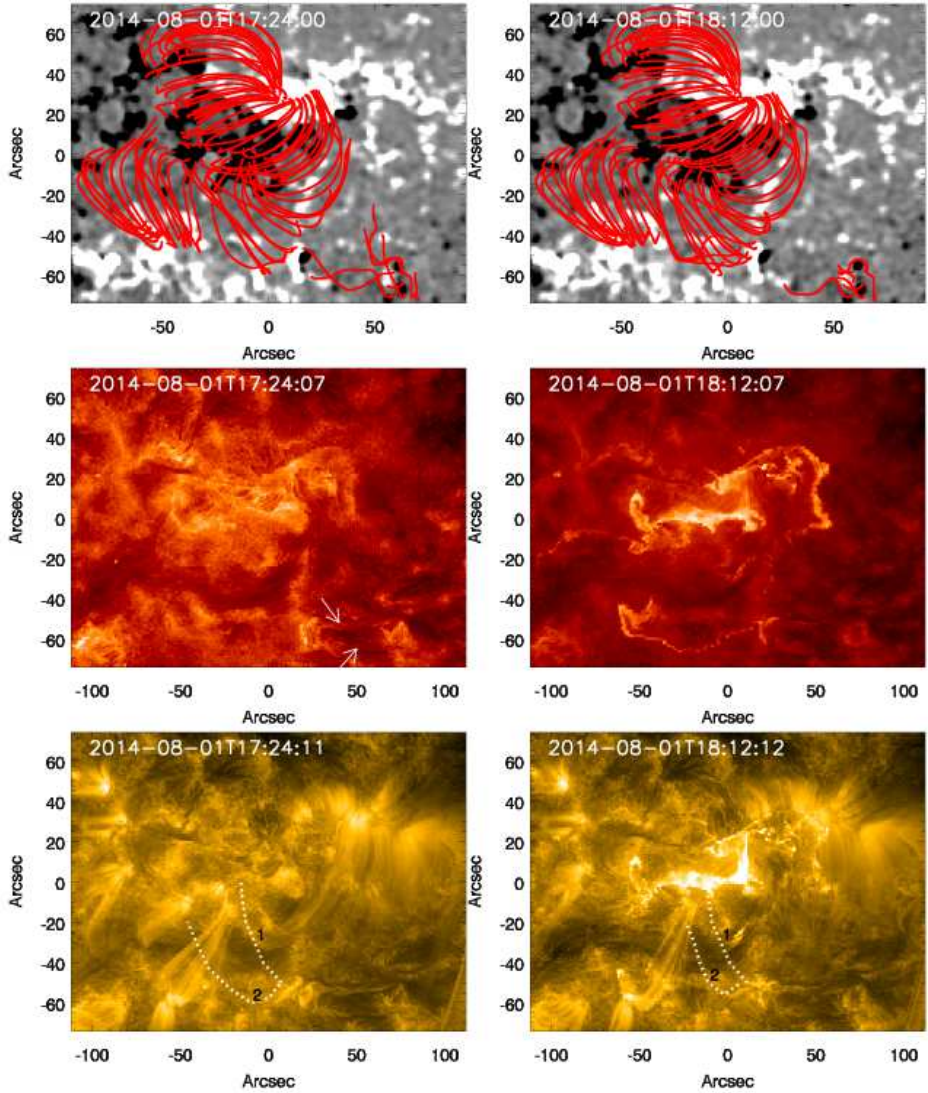}}
      \caption{Top: The distributions of magnetic-field lines obtained from the NLFF extrapolated field. Middle: AIA~304~\AA~images to show the lower layer observation compared with NLFF extrapolated results, especially for two small-scale magnetci structures below the main region, between them there exist magnetic connections and mass flows labeled by two white arrows in left panel. Bottom: AIA 171~\AA~images with dotted white lines indicate the instereted coronal loops that connect the main region and small scale magnetic structures. The left/right panel shows the results of NLFF field lines, 304~\AA~and 171~\AA~image before/after the flare.} \label{Fig4}
\end{figure}

To show the whole topology and properties of field-line connections for this active region, a non-linear force free (NLFF) extrapolation are used basing on vector magnetic field observation from HMI. Here the cadence of vector magnetic-field is 720 seconds and the NLFF extrapolation is optimization method
(\citeauthor{2004SoPh..219...87W}, \citeyear{2004SoPh..219...87W}). The distributions of magnetic-field lines from NLFF extrapolation shown in Figure \ref{Fig4}, where the 304~\AA~and 171~\AA~images are shown to compare the extents of match between extrapolated field lines and observations, such as coronal loops. The dots lines in the 171~\AA~image depict the interest coronal loops which connect
the main region and small-scale magnetic structure that probable trigger this M1.5 flare. On the whole, all NLFF extrapolated field lines can indicate the connectivities that observed in 171~\AA~ approximately, such as the connection between main region and the small magnetic structure can be found in the distribution of NLFF field lines.
In the Figure, two group images obtained before (left) and after (right) the flare are shown to tentatively find the possible changes in the magnetic field lines.
 It is found that there exist some low field lines between two small-scale magnetic structures, which also can be seen from 304~\AA~image where the connections and interactions labeled by two white arrows (the mass flow can be found in the 304~\AA~ movie) can be found, and this suggests that the lower layers may undergo low layers magnetic-reconnection processes. Then the large-scale field lines, such as those labeled in the 171~\AA~images, that connect the main region and the small magnetic structure should be changed consequently, due to the redistribution of these field lines. The above process may disturb the magnetic-field distributions of the whole active region and trigger this M1.5 flare.

\subsection{Magnetic Field Properties and Evolution}
\begin{figure}
   \centerline{\includegraphics[width=1\textwidth,clip=]{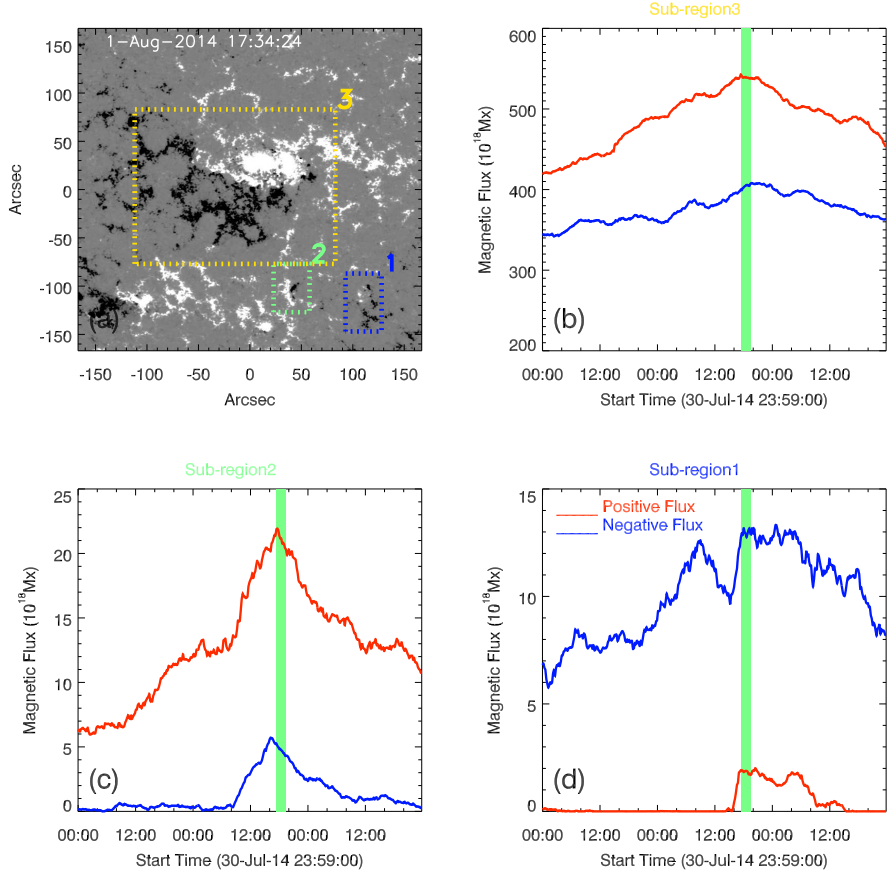}}
    \caption{(a): The LOS magnetic field with subregions labeled 1, 2, 3 and corresponding colors to show the evolution of magnetic flux in individual subregions. (b), (c) and (d): the evolution of magnetic flux for different subregions labeled in (a) with their number and color, where the green-vertical bars indicate the time when the flare was in progress. } \label{Fig5}
\end{figure}
Figure \ref{Fig5} shows the LOS magnetic field and the evolution of magnetic flux.
Similarly, there are three evident magnetic structures that can be identified easily; they are
indicated by subregions 1, 2, and 3, corresponding to those in the 304~\AA~images in Figure \ref{Fig3}.
At the mean time the evolution of magnetic flux corresponding to each of the subregion is shown and labeled in an individual.
The green bars plotted in each panel indicate the flare process.
From this figure, it can found that there exists a weak increase of magnetic flux for
subregion 3 (main flare region) before flare, followed by a decrease after the flare, and this trend is consistent for positive and negative fluxes.
As for subregion 2, it should be regarded as a long-lasting and slow process of magnetic emergence.
Where the small-scale magnetic structures with persistent magnetic emergences (SPME $\approx$ one day) connected main flare region, the properties of connection can be seen from 171~\AA~image in Figure \ref{Fig2} and NLFF extrapolated field lines in Figure \ref{Fig4}. During the process of emergence, especially for negative magnetic flux, the flux increases by a factor of 35 at its maximum.
As for the subregion 1, its magnetic flux has a quick increase near the beginning of flare, where small magnetic structures with impulsive magnetic emergences (SIME $\approx$ one to two hours) appear. For subregion 1, the positive magnetic flux of SIME increases by a factor of 20 during only about 1.5 hours, and the time of the maximum of SIME matches the flare time very exactly. So the SIME in subregion 1 probably plays a key role to trigger this main-region flare.
Meanwhile the effects of SIME on the flare operate through subregion 2 which connects to SIME and the main flare region.
Subregion 1, with evident and quick magnetic emergences, interacts with subregion 2 at first, and then the bright enhancements originating from
the magnetic reconnections and release of magnetic energy in subregion 2 are detected, which unavoidable affect and disrupt the
magnetic environments of subregion 3; finally the M1.5 flare is produced from the main region consequently.

\begin{figure}
   \centerline{\includegraphics[width=1\textwidth,clip=]{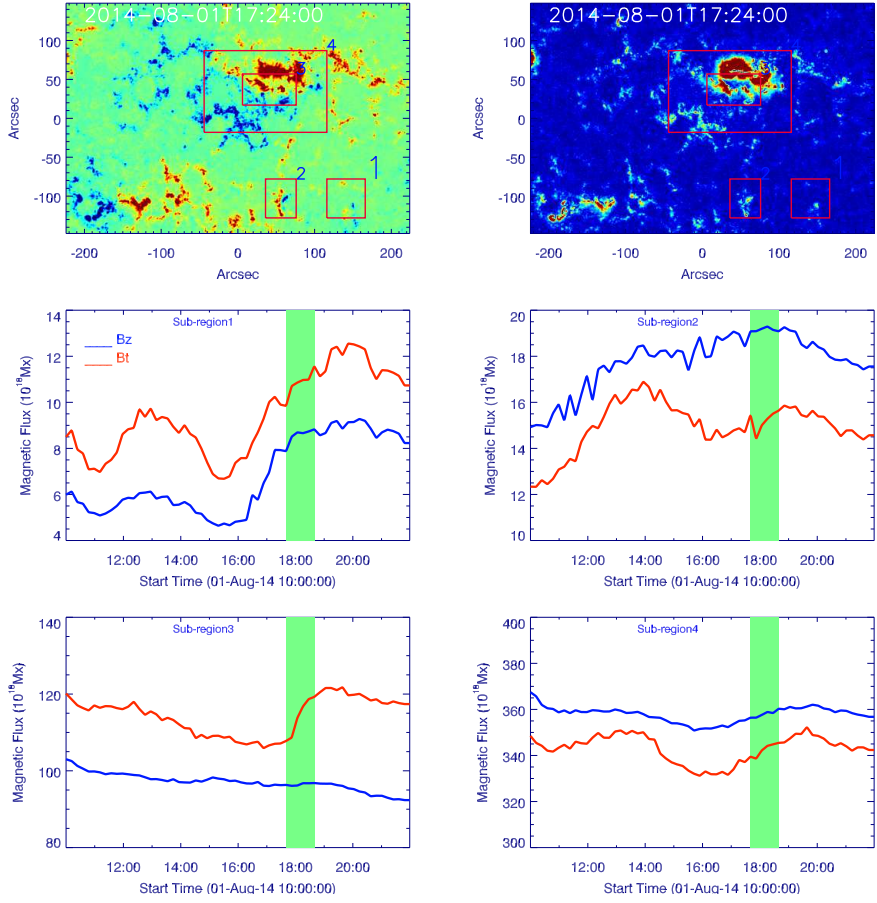}}
    \caption{The first row are LOS (left) and transverse (right) magnetic field at a given time, others are the evolution of LOS and transverse magnetic
    field for the subregion labeled in the first row, the green bars indicate the time when the flare was in progress, and the LOS and transverse field
    are drawn by blue and red lines, respectively. } \label{Fig6}
\end{figure}
Usually the magnetic field may have evident changes before and after the flare, as in Figure \ref{Fig6} the evolution of LOS and transverse magnetic field
are shown. To find the trend of changes, several regions are selected and indicated at the first row, where the left and right are LOS and transverse magnetic field, respectively.  The green-vertical bars indicate flare processes during evolution in the middle and bottom rows. Subregion 1 and 2 represent SPME and SIME, respectively. From their evolution, the trend of SPME with slow and long-lived magnetic field emergences
and SIME with a sharp magnetic field emergences are clearly seen before the flare erupted, while after flare the changes of magnetic field are not evident
for these two subregions. Subregion 3 indicates a region of which the main parts of magnetic neutral line are contained; it can be found that the transverse magnetic field
has an evident sharp enhancement after the flare erupted, which means magnetic field lines located the near neutral line tend to horizontal due to the flare.
Subregion 4 contains main magnetic-field region related to the flare, for the transverse magnetic field there are weak trends of enhancements after the flare.
The above results are consistent with previous studies that enhancements of transverse magnetic field were found after a flare for some active regions; this phenomenon is very evident especially for the interesting region dominated by neutral lines
(\citeauthor{2005ApJ...627.1031W}, \citeyear{2005ApJ...627.1031W};
\citeauthor{2011ApJ...733...94S}, \citeyear{2011ApJ...733...94S};
\citeauthor{2012ApJ...745L..17W}, \citeyear{2012ApJ...745L..17W}),
and this situation matches the theoretical models (\citeauthor{2008ASPC..383..221H}, \citeyear{2008ASPC..383..221H}).

\subsection{Sunspot Rotation and Helicity}
\begin{figure}
   \centerline{\includegraphics[width=1\textwidth,clip=]{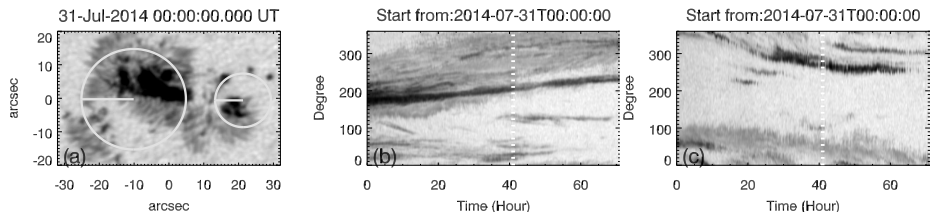}}
    \caption{Sunspot rotation at main region of flare displayed by continuum observations; the two white circles drawn in the left panel, of which starting position (0 degree) indicated by corresponding white lines, and the angles increase through counter-clockwise rotation. The middle panel shows the big circle rotations by time--distance plots, while the right panel corresponds to the small circle. The doted lines drawn in the middle and right panel indicates the time when the flare was in progress.} \label{Fig7}
\end{figure}
This M1.5 flare is triggered by complex effects that originate from the overall effects of subregions 1 and 2, while the released energy should be contained in the main region.
So, we pay close attention to investigating the properties of main region related flare. Based on LOS magnetic field and continuum series observations, it can be found that there exist evident large-scale
rotations at positive magnetic field of main region (subregion 3 in Figure \ref{Fig5}). In Figure \ref{Fig7}, through the time--distance plots, the rotations of sunspots are shown qualitatively and quantitatively. In the left panel of Figure \ref{Fig7} two circles with evident rotations are selected; the initiation position labeled by the white horizontal line with definition of 0 degree and the increase of angle is indicated by counter-clockwise rotation. Then the circles in the time series of observed images are spread in cartesian coordinates; the middle and right panels correspond to the big and small circle, respectively. Through the evolution of characteristic structure in these circles, the rotations of the sunspot can be distinguished easily. Through the characteristic structure recognized and calculated in the cartesian coordinate system, it is found that the big circle has the rotation velocity of counter-clockwise 1.38$^\circ$~hour$^{-1}$ and the small circle is clockwise 1.52$^\circ$~hour$^{-1}$ before the flare, while after the flare the rotation of small circle decreases. For NOAA 12127, there are only four flares with C-class which all occur
40 hours before this M1.5 flare (three flares erupted within 50 hours); except for this M1.5 flare the active region remains relatively quiet. However, when the active region is triggered by small
magnetic emergences M-class flare is produced; the rotations of the sunspot assuredly contribute to the accumulation of energy (after this M1.5 flare there are no other flares for this active region).

\begin{figure}
   \centerline{\includegraphics[width=1\textwidth,clip=]{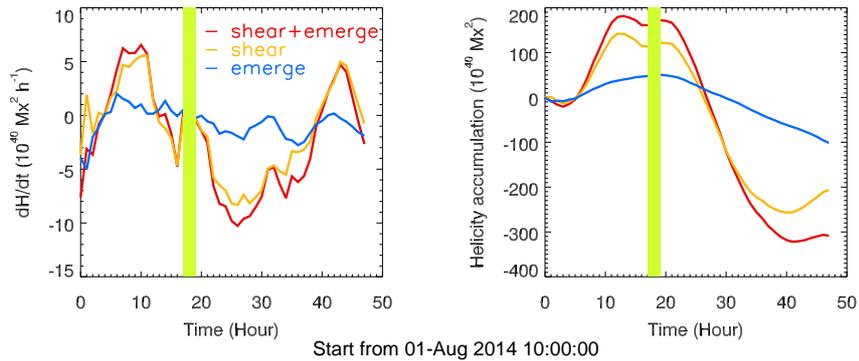}}

    \caption{The evolutions of helicity-injection rate and accumulation calculated by \textbf{DAVE4VM} the method, left panel shows the helicity-injection rate, while the right is the helicity accumulation calculated. Here the helicity produced by magnetic shear and emergence are calculated individually, and the results are shown by corresponding color curves. The vertical color bar in each panel indicates time when the flare was in progress.} \label{Fig8}
\end{figure}

Figure \ref{Fig8} shows the helicity-injection rate and helicity accumulation calculated by the differential affine velocity estimator for vector magnetograms (\textsf{DAVE4VM};
\citeauthor{1984JFM...147..133B}, \citeyear{1984JFM...147..133B};
\citeauthor{2008ApJ...683.1134S}, \citeyear{2008ApJ...683.1134S}). The helicity created by magnetic shear and emergence can be deduced individually when the \textsf{DAVE4VM} method is used, hence these two parts can be studied and compared. In Figure \ref{Fig8} the helicity-injection rate and helicity accumulation originated from magnetic shear, magnetic emergence, and both of them (shear+emergence) are shown by different color curves.
From this figure, it is found that the helicity is mainly produced by magnetic shear for this active region; there exist differences in the order of helicity magnitude created by magnetic shear and emergence,
so the total of  helicity injection rate and helicity accumulations are basically the same as those created by magnetic shear (red and yellow lines in two panels), which also means that sunspot rotation  should certainly be the main contribution to the creation of magnetic helicity.
Before flare the positive helicity injection is dominant, hence the positive helicity accumulation reach its maximum before the flare erupted, and the helicity accumulations decrease due to negative helicity injection. There are evident watersheds at the flare time both for helicity injection and accumulations, which means that helicity accumulation with the same signs probably contribute to the accumulation of energy for this M1.5 flare. The results from the calculation of helicity are very similar to those of \citeauthor{2008ApJ...686.1397P} (\citeyear{2008ApJ...686.1397P}) for a fraction of active regions.
\citeauthor{2008ApJ...686.1397P} (\citeyear{2008ApJ...686.1397P}) found a phase of monotonically increasing helicity and a following phase of relatively constant helicity before a flare.
The event studied in this work also shows an evident change of helicity trend before and after the flare. Additionally, the main contribution from magnetic shear motion (sunspot rotations) to helicity can be determined by the \textsf{DAVE4VM} method.

\section{Discussion and Conclusions}
In this article, a M1.5 flare process is studied through magnetic field and continuum observed by HMI and multi-band AIA observations.
This event can be regarded as a high-quality example to study the processes of a flare, because there are only four C-class flares that occur before this M1.5 flare.
Additionally, except for a flare that erupts 40 hours before this M1.5 flare, the others three flares occur at least 50 hours before, so the confusing effects of multiple flares are avoided. Triggering mechanism, magnetic-field evolution, the rotation of the sunspot related to the flare are regarded as the main aspects to be studied.

Before this M1.5 flare, the active region remains quiet for more than 40 hours, and during this time there appears a SPME that connects to the main active region.
Furthermore, SPME interacts with another small-scale magnetic structure (SIME), and between them (SPME and SIME) there exist intermittent mass flows.
Near the time of flare eruption, the emergence of SIME become very remarkable, and the brightness of SIME begins to be clearly enhanced.
Consequently, bright enhancements of SPME become very noticeable (these bright enhancements can be regarded as a very small flare activity), and at last due to the triggers of SPME the flare erupts.
The time of SIME and flare match very well, so SIME plays a key role to produce the flare. While the SPME plays a bridge action between flare and SIME,
the trigger mechanism and the associated chain reactions become major factors to initiate this M1.5 flare.
\begin{figure}
   \centerline{\includegraphics[width=0.85\textwidth,clip=]{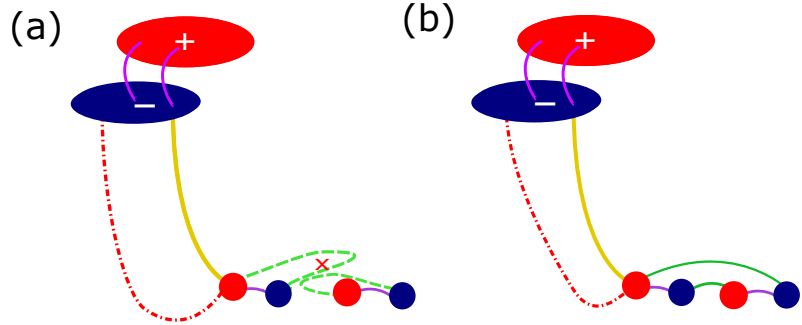}}
    \caption{ The model possibly explains the magnetic reconnection and flare trigger processes; in each panel the upper parts indicate main region with large-scale positive and negative magnetic field labeled by red and blue ellipses and the lower parts correspond SPME and SIME, respectively. The broken field lines are drawn as the green dashed line (a) with the red crosses showing the possible position of disconnection, and the corresponding reconnected lines are indicated by solid green lines (b). The red and yellow lines in the low parts are the large-scale coronal loops that connect the main region and small-scale magnetic structures; these lines play the key role to trigger the M1.5 flare. The purple lines are ordinary magnetic-file lines connecting positive and negative polar.
 Here only the interesting field lines are shown, which take part in the main processes of the flare.} \label{Figmodel}
\end{figure}

\begin{figure}
   \centerline{\includegraphics[width=1\textwidth,clip=]{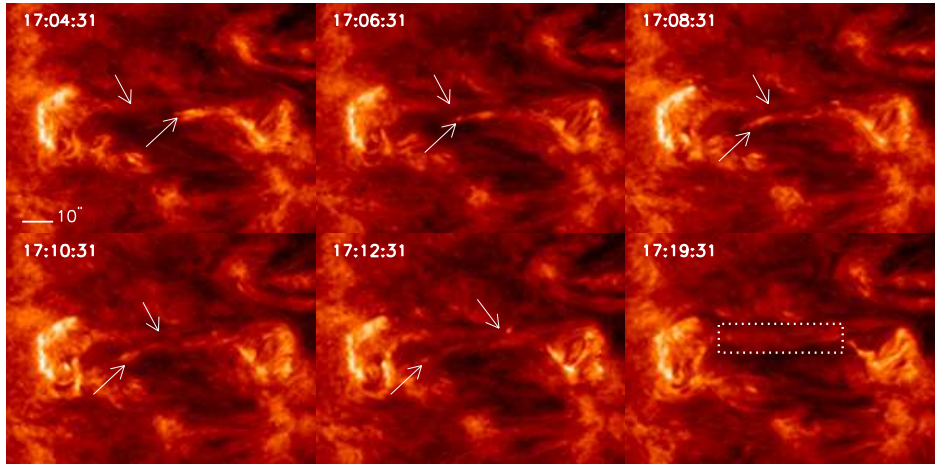}}
    \caption{An example to show dynamic processes observed in  AIA 304~\AA~channel; the white arrows indicate the evolution of this dynamic process; the white rectangle are the last state of this process.} \label{Figsmall304}
\end{figure}

Figure \ref{Figmodel} shows a cartoon that depicts the possible process of magnetic reconnection and flare trigger.  In each panel, the upper parts indicate the main regions with large-scale positive and negative magnetic field labeled by red and blue ellipses, the lower parts correspond to SPME and SIME, respectively. Here the green dashed (panel a)/solid (panel b) lines indicate the broken/reconnected field lines during reconnection processes.  The red and yellow lines in the lower parts are the large-scale coronal loops that connect main region and small-scale magnetic structures (SPME). The red cross (panel a)  indicates the possible position of broken field lines.
It shows the small-scale magnetic reconnections happen between two small-scale magnetic structures in the lower parts, hence the energy releases and bright enhancements are detected at the position of SPME. Due to their small-scale, the energy releases are not too large, but the bright enhancements are very evident. These processes should be regarded as the low-layer magnetic reconnection that models and matches the 304~\AA~observation, where evident and intense magnetic activities occur (two white arrows drawn in Figure \ref{Fig4}: left middle panel ). To see the magnetic activity more intuitively, oen example of times series of 304~\AA~observations are shown by small field-of-view in Figure \ref{Figsmall304}. Here we can see the dynamic process labeled by white arrows and rectangle in the 304~\AA~observations, which may suggest
that there exist magnetic reconnections in the lower layer, and this process is depicted by the description of possible  magnetic reconnections shown in the cartoon.
In Figure \ref{Figsmall304}, the structures marked by the two white arrows correspond to the green dotted lines in panel a of Figure \ref{Figmodel}, the features enclosed by rectangle correspond the green field lines in panel b of Figure \ref{Figmodel}, and the possible magnetic reconnections are simulated by green dotted/solid lines in panel a,b of the cartoons.
The above small-scale magnetic reconnections are regarded as the trigger of main process of M1.5 flare. Naturally, large-scale coronal loops take an important role in this processes. It is noted that the red line undergo evident changes of shape, which imitate the loop 2 labeled in Figure \ref{Fig4}. The reason for this change should be some space magnetic environmental disturbance. While the changes of the yellow line are not evident (or the changes are too weak to be distinguished), hence the exact situations and effects of yellow lines are not explained in depth, but the magnetic-field disturbance can also be transferred to main region through these yellow filed lines. It should note that the trigger of small-scale magnetic reconnections eventually cause the high intensity magnetic reconnection inside the main region to cause the eruption of the M1.5 flare. In short, it should be that the comprehensive effects resulting from the interactions between SPME and SIME trigger this M1.5 flare.

\citeauthor{2017NatAs...1E..85W} (\citeyear{2017NatAs...1E..85W}) studied a precursor of this flare using high-resolution observations obtained at the Big Bear Solar Observatory; they confirmed that small-scale bright enhancements inside an active region should be regarded as the precursor of its following solar flare. The results of  \citeauthor{2017NatAs...1E..85W} (\citeyear{2017NatAs...1E..85W})  also provide evidence of low-atmospheric small-scale energy release, possibly related to the onset of the main flare. The trigger of flare in this M1.5 flare should be
also small-scale magnetic structure emergence, which are more complex small-scale magnetic activities. However the small-scale brightenings (magnetic structures) are not inside the active region as described by \citeauthor{2017NatAs...1E..85W} (\citeyear{2017NatAs...1E..85W}), but they ar remote bright enhancements. Hence the large-scale coronal loops that connect the main region and remote bright enhancements play a bridge role, which can be seen from NLFF extrapolated field lines and corona observations in Figure \ref{Fig4}.

The M1.5 flare erupted from this relatively quiet active region suddenly,; the accumulation of magnetic energy is necessary for this relatively big
flare with approximately closed circular ribbons. Through the \textsf{DAVE4VM} method the magnetic helicity-injection and accumulations are calculated, and it is found that
before the flare there are evident magnetic helicity injection to the corona; consequently there are accumulations of magnetic helicity and energy.
For this magnetic-helicity injection, namely magnetic-energy accumulations, the magnetic shear motions contribute to the most of parts of helicity.
Hence, the large-scale rotations of the sunspot at the main flare region certainly play key roles.

\clearpage
\begin{acks}
We thank referee for valuable suggestions and constructive criticism
which improved the clarity of the article. This work was partly supported by the Grants:  KJCX2-EW-T07,
2014FY120300, 11203036, 11473039,  U1531247, 11673033, 11773038 and
11703042.
The Strategic Priority Research Program on Space Science, CAS,
Grant No. \\ XDA15052200 and XDA15320201. The
Key Laboratory of Solar Activity National Astronomical Observations, Chinese Academy
of Sciences.

\end{acks}

\section*{Disclosure of Potential Conflicts of Interest}
The authors declare that they have no conflicts of interest.



\end{article}

\end{document}